\documentstyle[12pt,amssymb]{article}
\def\BZ{\mbox{$\Bbb Z$}} \def\BR{\mbox{$\Bbb R$}}
\def\BC{\mbox{$\Bbb C$}} \def\BP{\mbox{$\Bbb P$}}
\def\CP{\BC\BP}

\begin{document}
\begin{titlepage}
\renewcommand{\thefootnote}{\fnsymbol{footnote}}
\begin{flushright}
\vbox{\begin{tabular}{l}\\
{\tt BONN-TH-97-08}\\
{\tt hep-th/9707164}\\
July 28, 1997
\end{tabular}}
\end{flushright}
\begin{center}
{\large {\bf Heterotic M(atrix) theory at generic points in Narain
moduli space}}
\end{center}
\bigskip
\centerline{\it Suresh Govindarajan\footnote{On leave from the
Department of Physics, Indian Institute of Technology, Madras 600 036,
India.}}
\centerline{\it Physikalisches Institut der Universit\"at Bonn}
\centerline{\it Nu\ss allee 12, D-53115 Bonn, Germany }
\centerline{\tt Email: suresh@avzw02.physik.uni-bonn.de}
\begin{abstract}
Type II compactifications with varying string coupling can be described
elegantly in  F-theory/M-theory as compactifications on U - manifolds.
Using a similar approach to describe Super Yang-Mills with a varying coupling
constant, we argue that at generic points in Narain moduli space, the $E_8
\times E_8$ Heterotic string compactified on $T^2$ is described in M(atrix)
theory by ${\cal N}=4$ SYM in 3+1 dimensions with base $S^1 \times \BC\BP^1$
and a holomorphically varying coupling constant.  The $\BC\BP^1$ is best
described as the base of an elliptic K3 whose fibre is the complexified
coupling constant of the Super Yang-Mills theory leading to manifest U-duality. 
We also consider the cases of the Heterotic string on $S^1$ and $T^3$.
The twisted sector seems to (almost)
naturally appear at precisely those points where enhancement of
gauge symmetry is expected  and need not be postulated.
A unifying picture emerges in which the U-manifolds which 
describe type II orientifolds  (dual to the Heterotic string) as M- or F-
theory compactifications play a crucial role in Heterotic M(atrix)
theory compactifications.
\end{abstract} 
\end{titlepage}

\section{Introduction}

Our improved understanding of duality symmetries, both perturbative and
non-perturbative, suggests that the five string theories as well as eleven
dimensional supergravity are limits of a single theory, an eleven
dimensional theory called M-theory.  Some insight into this theory has been
provided by looking at strong coupling limits of various string
theories\cite{strong}.  A new tool in understanding M-theory was provided by
Banks, Fischler, Shenker and Susskind\cite{bfss} who introduced a M(atrix)
theory which was 
obtained by considering the strong coupling limit of N D0-branes in
IIA string theory.  The quantum mechanics of this M(atrix) theory\cite{qall}
 was proposed as a candidate for M-theory in the Infinite Momentum Frame
(IMF) with `N', the number of D0-branes being identified with the eleven
dimensional momentum. This theory has passed a number of tests.
We refer the reader to a recent review\cite{review}
for a description of the details of those results.

Further
compactifications of M(atrix) theory on tori $T^d$ have been useful
in understanding aspects of this theory.  For compactifications where
$d\leq3$, this leads to Supersymmetric Yang-Mills (SYM) in
$d+1$ dimensions with base manifold given by the dual torus
$\widetilde{T}^d$.\cite{comp,sdual,fhrs}  More importantly, 
U-duality\footnote{This is a
conjectured discrete symmetry which includes all perturbative and
non-perturbative symmetries of string theory\cite{ht}.} is visible in
M(atrix) theory with excited states sitting in representations of the
appropriate U-duality group.  For $d>3$, the SYM prescription breaks down. 
This breakdown can be seen in two ways: The full U-duality group is not
realised and these are non-renormalisable quantum field theories.  Both
problems suggest the need for additional degrees of freedom to make the SYM
prescription complete.  This has led to the construction of new theories
which exhibit manifest U-duality\cite{four,five}.  One important consequence
is that the base manifold of M(atrix) theory is related to the target
spacetime only in special limits.  For example, even a simple topological
quantity like the dimension of the (compact part of) target spacetime and
that of the base manifold may not be the same. 

Another class of interesting compactifications of M(atrix) theory are those
which involve lesser supersymmetry.  The simplest class of these models is
Heterotic M(atrix) theory and its compactifications\cite{het,petr}.  Other
examples are compactification of M-theory on K3, IIB on K3, F-theory
compactified on elliptic K3 as well as non-compact examples such as ALE spaces
and non-compact orbifolds\cite{noncomp}.  These theories all have eight
unbroken supercharges and hence supersymmetry imposes fewer constraints
here.  Thus these models are still not as well understood as the case with
sixteen unbroken supercharges except at special points in their moduli
spaces.  However, recently there has been progress for the case of Heterotic
M(atrix) theory on $S^1$\cite{krey}.  It has also been noticed that some of
the known non-perturbative dualities have interesting realisations in
M(atrix) theory.  For example, M-theory on K3 and Heterotic string on $T^3$
are different limits of the same M(atrix) theory\cite{seven,BZ,hrey}. 

Even though several of our considerations will be rather general,
the main focus of this paper will be the case of Heterotic M(atrix) on
$T^2$. We will also have some remarks for the case of compactification
on $S^1$ which has been recently studied by Kabat and
Rey\cite{krey} and for the case of $T^3$ compactification. 
The moduli space of Heterotic string theory compactified
on $T^d$ is given by the Narain moduli space\cite{narain}
\begin{equation}
\Gamma_d \backslash {{SO(16+d,d)}\over{SO(16+d)\times SO(d)}} \times \BR^+
\quad,
\end{equation}
where $\BR^+$ corresponds to the heterotic dilaton and $\Gamma_d =
SO(16+d,d;\BZ)$ is the U-duality group for these theories(for $d\leq3$).  
For $d=1$, there
are 18 real moduli and for $d=2$, there are 18 complex moduli (excluding the
dilaton).  Of these, 16 complex (real) moduli correspond to the Wilson lines of
$E_8 \times E_8$ for d=2 (d=1).  A complete description of the Heterotic
M(atrix) compactified on $T^d$ should include a manifest realisation of this
symmetry.  It is also clear that such a description will include
generic points in  Narain moduli space.  

The working prescription for describing the Heterotic string on $T^d$ in
M(atrix) theory is to consider the orbifold U(N) SYM in $d+2$ dimensions
with base $\widetilde{S}^1\times (\widetilde{T}^d/\BZ_2)$\cite{het,petr}. 
Anomaly
cancellation is accomplished by including a twisted sector given by 32
fermions distributed equally among the $2^d$ fixed points (circles to be
precise) of the orbifold $\BZ_2$.  This choice corresponds to a point in
Narain moduli space with enhanced gauge symmetry $[SO(32/r)]^r$, where
$r=2^d$.  Since the prescription assumes the U(N) SYM description for
M(atrix) theory on $T^{d+1}$, we will restrict to the cases $d\leq2$
though we will have some remarks for $d=3$. 
\footnote{The cases of $d=3,4$ can in principle be included by using the
orbifolding procedure on the theories proposed for M(atrix) theory on $T^4$
and $T^5$\cite{four,five}.This will however require a more detailed
understanding of these new theories.}

The problem of describing M(atrix) theory at more generic points in 
Narain moduli space can be described in two ways: The first is to view it as
a clear description  of Wilson lines in Heterotic M(atrix) theory.  The second
is to view it as providing a manifest U-dual description of Heterotic M(atrix)
theory.  These are complementary views and hence understanding one should
shed light on the other.  This paper pursues the second perspective.  An
important observation (see Sec. 4 for details)
to make here is that for both $d=1,2$ (as well as for
other values of $d<5$), the Heterotic string at the point in moduli space
where the enhanced gauge symmetry is $[SO(32/r)]^r$, the theory is dual to
orientifolds of type II string theory (IIA for odd $d$ and IIB for even
$d$)\cite{porient}.  Further, this duality maps Wilson lines to the positions
of certain D-branes which have to be included into the orientifolding for
cancelling RR charges located at the fixed points of the orientifold group. 
In these cases, generic points in the moduli space correspond to moving the
D-branes away from the orientifold fixed point.  However, this involves
considering moduli which depend on the compactified coordinates and includes
regions where one is at strong coupling\cite{polwit,ashoke}. 

How does one view varying spacetime moduli in M(atrix) theory.
Generically, one expects that these moduli will enter as parameters in
the base manifold. On D-brane probes, one sees however that varying
moduli (such as the dilaton) lead to a varying coupling constant in the
worldvolume theory.  {\it We shall make the ansatz that varying moduli in
spacetime should correspond to introducing varying
coupling constants in M(atrix) theory}. This was first proposed
by Kabat and Rey in the context of Heterotic M(atrix) theory on 
$S^1$\cite{krey}. 
Thus our problem can be posed as the need to understand the situation of SYM
with a varying coupling constant such that U-duality is manifest with 
the appropriate amount of supersymmetry being broken. 

This might seem to be a rather difficult problem to solve.  We solve this 
by a method used in F-theory\cite{vafa} by considering a $d+s$ dimensional 
fibred manifold ${\cal Y}$
(we will refer to this manifold as  the U-manifold following \cite{alok})
with base ${\cal B}$ of dimension $d$ and fibre is identified with the
coupling constant of the SYM theory.  Now consider compactifying the 
M(atrix) theory which describes M-theory on $T^r$ (for some $r>d+1$)
 and compactify it on a base $S^1\times
{\cal Y}$.  In the limit of large base and non-singular fibres, this can be 
viewed as SYM with base
$S^1 \times {\cal B}$ and a coupling constant varying with respect to the
coordinates on ${\cal B}$ as dictated by the geometry of ${\cal Y}$.  An
important constraint on ${\cal Y}$ is that there must be a special point in
its moduli space where ${\cal B}$ can be identified with
$T^d/\BZ_2$ where the fibre is constant.  At this point, we recover the
standard prescription of SYM on $S^1\times (T^d/\BZ_2)$\cite{het,petr}.  Other
conditions on ${\cal Y}$ are that it break half the supersymmetry and have a
moduli space which can be identified with the Narain moduli space.  Though
the above is probably 
true in general we will discuss the candidates for $d=1,2,3$ in
this paper as concrete examples. In these cases, we will see that the
results seem consistent with existing proposals for Heterotic M(atrix)
theory based on known string dualities\cite{seven,BZ,hrey}. Further, we
will see that the twisted sector emerges without being postulated at 
precisely the points where enhanced gauge symmetry occurs (though we do
not understand the $d=1$ case very well).

The plan of this paper is as follows. In Sec. 2, we review some of the
relevant aspects of Heterotic M(atrix) theory on Tori. In Sec. 3 we
discuss some aspects of U-duality for toroidal compactifications
in M(atrix) theory.
We consider in some detail the seemingly trivial example of M(atrix)
theory on $T^3$ where the
condition of geometrising U-duality is analogous to the relation between 
the IIB string and F-theory. This provides the motivation for the
approach pursued in this paper. In Sec. 4, we discuss three orientifolds
of type II theory which are dual to the $E_8\times E_8$ Heterotic string
compactified on Tori. We show how these orientifold compactifications
of type II theory can be best described as M- or F- theory
compactifications on U-manifolds. In Sec. 5, we come to the main result
of the paper. Here we show how in Heterotic M(atrix) theory away from
its orbifold limit can be described by using the same U-manifolds which
occurred in Sec. 4. We also relate to existing proposal for Heterotic
M(atrix) theory. In Sec. 6 we end with some concluding remarks. An
appendix is devoted to some relevant details of F-theory compactified on 
an elliptic K3.

\section{Heterotic M(atrix) theory on Tori}

We shall collect some of the relevant details of Heterotic M(atrix)
theory on tori. See \cite{het,petr} for more details. The Lagrangian for
M(atrix) theory on $T^d$ (for $d<4$) is given by U(N) SYM with 16 unbroken
supercharges in $d+1$ dimensions. This can be obtained from the
dimensional reduction of the ten dimensional SYM with ${\cal N}=1$
supersymmetry. The bosonic part of the Lagrangian is given by
\begin{equation}
L_b = {1\over {g^2}} \int_{\widetilde{T}^d\times R} d^d\sigma dt\ tr \left \{
-{1\over4} F^2 + {1\over2} (D_\alpha X^i)^2 -{1\over4} ([X^i,X^j])^2
\right \} \quad,
\end{equation}
where $X^i$ for $i=1,\ldots,(9-d)$ are adjoint valued scalars and
$D_\alpha$ is the covariant derivative with $\alpha=0,\ldots,d$. 
The base of the SYM is indicated by $\widetilde{T}^d$.\footnote{We shall
indicate the dual tori with a tilde in most cases. Since there are two
different but related manifolds in the picture: we shall refer to the
one in physical spacetime as the target manifold and the one associated
with SYM as the base manifold.} When the torus
is rectangular, its sides $\Sigma_\alpha$ are  related to that of the
target space torus $T^d$, $R_\alpha$ by
\begin{equation}
\Sigma_\alpha \sim 1/(r R_\alpha)\quad,
\end{equation} 
in units where $l_{11}=1$ (and ignoring constants). The eleven
dimensional momentum $p_{11}=N/r$ and $r$ is the radius of the
eleventh dimension which emerges in the strong coupling limit as
argued in ref. \cite{bfss}.
The SYM coupling $g$ is related to the spacetime parameters by
\begin{equation}
g^2 \sim {r^{3-d} \over V}\quad,
\end{equation}
where $V$ is the volume of $T^d$. 

The Heterotic string on $T^{d-1}$  is obtained in M-theory as the
orbifold $(S^1)/\BZ_2 \times T^{d-1}$\cite{hw}. 
The orbifolding is implemented in M(atrix) theory by appropriately
embedding the $\BZ_2$ in the U(N) SYM in $d+1$ dimensions. (We closely
follow the ref. \cite{petr} here.)
The result of this orbifolding is that Heterotic M(atrix) theory
is described by SYM on base ${\cal B}=\widetilde{S}^1 \times
(\widetilde{T}^{d-1}/\BZ_2)$. 
The bosonic part of the Lagrangian is given by
\begin{equation}
L_b = {1\over {g^2}} \int_{{\cal B}\times R} dt d^d\sigma\ tr \left \{
-{1\over4} F^2 + {1\over2} (D_\alpha X^i)^2 -{1\over4} ([X^i,X^j])^2
\right \} \quad,
\end{equation}
with the sides $\Sigma_\alpha$ and SYM coupling constant as given
before. ($\alpha=1$ will represent the orbifold direction in target
space.) At the $r\equiv 2^{d-1}$ fixed circles of the
orbifold group, the U(N) gauge symmetry is broken to $O(N)$ with half
of the supersymmetry being broken. The theory as we described it
is anomalous. As in any orbifolding with fixed points, one includes a
twisted sector localised at the fixed points. The twisted sector 
is provided by 32 fermions transforming in the fundamental
representation of $O(N)$ which
seems to be (almost) uniquely determined by anomaly cancellation. 
(These are equivalent to 16 chiral bosons in $1+1$ dimensions after
bosonisation.) They are described by the action
\begin{equation}
\sum_{i=1}^{32} \int dt d\sigma^1\  (\chi_i^I\ (D_+)^{IJ}\ \chi_i^J )\quad,
\end{equation}
where $\chi_i$ are the fermions and $D_+$ is the covariant derivative.
Local anomaly cancellation at each of the fixed circles seems to suggest
that the 32 fermions be distributed equally among the $r$ circles
(this is only possible for $d<5$).  We will assume that this is the
case. Then one obtains a global symmetry $SO(32/r)$ at each of the fixed
circles. This implies a gauge symmetry in target spacetime following the
rule that global symmetry on the D-branes should be interpreted as gauge
symmetries in target space. Ho\v rava has provided a nice description of
the situation where the fermions are not localised at the fixed circles
using anomaly inflow arguments\cite{petr}. A more detailed analysis
has been done by Kabat and Rey for the $S^1$ case where they find a
relationship between the gauge coupling and Chern-Simons term (which
they introduce) based on {\it local} anomaly cancellation\cite{krey}.
In addition, as the fermions are moved away from the fixed points, they
pick up masses (for the same reason that open strings connecting separated
D-branes pick up a mass.) 

\section{Manifest U-duality in M(atrix) theory}

As mentioned in the introduction the key success of M(atrix) theory is in
realising U-duality as a geometric symmetry.  For M(atrix) theory
compactified on $T^2$, the U-duality group is $SL(2,\BZ)$ which is the
mapping class group of the base two torus of SYM.  M(atrix) theory on $T^4$
is described by a $(2,0)$ theory in 5+1 dimensions compactified with base
$\widetilde{T}^5$.  The U-duality group $SL(5,\BZ)$ is the mapping class
group of the base torus.  For the case of $T^5$, the U-duality group
$SO(5,5;\BZ)$ is the T-duality of a six dimensional string theory
compactified on a base $\widetilde{T}^5$.  Thus these cases all realise
U-duality as a geometric symmetry of the underlying field/string theory.  It
is still an open problem to understand the case of M(atrix) theory on $T^d$
for $d>5$. 

We will now consider the case of $d=3$ not discussed above.  This is
an interesting situation which lies at the edge of where the SYM
prescription breaks down.  The U-duality group here is $SL(3,\BZ)\times
SL(2,\BZ)$ and M(atrix) theory is described by ${\cal N}=4$ SYM in 3+1
dimensions with base $\widetilde{T}^3$.  The $SL(3,\BZ)$ is geometrically
realised as the mapping class group of the base torus while the $SL(2,\BZ)$
corresponds to the electric-magnetic duality which is a quantum
non-perturbative symmetry of ${\cal N}=4$ SYM in 3+1 dimensions.  However,
as is clear, this description has one shortcoming: the full U-duality
group is not realised as a geometric symmetry.  This is similar to the case
of IIB string theory in ten dimensions which has a $SL(2,\BZ)$ symmetry. 

For the IIB string, Vafa proposed F-theory in order to geometrically realise
the $SL(2,\BZ)$\cite{vafa}. F-theory can be considered to be a IIB
compactification where the dilaton and RR scalar are not constant, but
vary with respect to the internal manifold. Given a manifold ${\cal Y}$
which is elliptically fibred with base ${\cal B}$, 
F-theory on ${\cal Y}$ can be defined to be IIB compactified on ${\cal
B}$ with the dilaton/RR
scalar being set equal to the complex structure of the fibre torus,
The $SL(2,\BZ)$ of IIB theory is the geometric symmetry of the torus.
For the ten dimensional IIB string this is a simple way to
geometrically view the $SL(2,\BZ)$ as acting on some auxiliary torus.
(Of course, it is possible that this theory might exist as a 12
dimensional theory.) A non-trivial example is furnished by
considering F-theory on an elliptic K3. This becomes a consistent 
compactification
provided one includes 24 D7 branes and has been shown to be dual to the $E_8
\times E_8$ Heterotic string on a two torus. 

Similar considerations can
be extended to IIA theory which can be viewed as M-theory compactified
on a circle. Non-trivial IIA compactifications can be obtained by
considering a fibred manifold ${\cal Y}$ with base ${\cal B}$ and
fibre $S^1$. As in F-theory, M-theory on ${\cal Y}$ can be viewed as
IIA on ${\cal B}$ with the string coupling  related to the radius of the fibre.
We shall refer to these manifold on which F- and M- theory are
compactified as U-manifolds following ref. \cite{alok} since U-duality
is manifest as a geometric symmetry of these manifolds.

Inspired by F-theory,  we return to the case of M(atrix) theory on $T^3$, 
where we propose to introduce an auxiliary torus whose complex
structure is the complexified coupling constant of the SYM theory.  This is
also motivated by the work of Verlinde who discussed electric-magnetic duality
for U(1) SYM in 3+1 dimensions\cite{verlinde}.  One begins with  a 5+1
dimensional theory with a self-dual two-form gauge field which reduces to
the $U(1)$ gauge field and its magnetic dual on compactifying on an
auxiliary torus.  In the supersymmetric context, the self-dual two-form
gauge field becomes part of a tensor multiplet which is the only matter
multiplet in $(2,0)$ theories in 5+1 dimensions.  For the $U(N)$ case, the
appropriate theory is the same $(2,0)$ theory proposed in ref.  \cite{four}
to describe M(atrix) theory on $T^4$!  
Thus requiring manifest $SL(2,\BZ)$
implies that we should consider the $(2,0)$ theory proposed in ref.
\cite{four}
compactified on `$T^2$'$\times \widetilde{T}^3$ where `$T^2$' is an
auxiliary torus on which electric-magnetic duality is geometrically realised
and $\widetilde{T}^3$ is the base of SYM.\footnote{This observation has also
been made by E.  Verlinde in his seminar at Strings '97.}

One might wonder as to why we seem to have considered this result in so much
detail.  The reason is that just as M-/F-theory provides a large number of
non-trivial examples of type II compactifications with lesser supersymmetry, 
this picture will prove useful in understanding M(atrix) theories with 
lesser supersymmetry. As it turns out the same U-manifolds which occur
in M-theory and F-theory turn up in Heterotic M(atrix) theory!

\section{Orientifolds in type II string theory}\label{orient}

In this section, we will consider some relevant details of the theories dual
to the heterotic string on $T^d$ for $d=1,2,3$.  These dual theories have a
special point in their moduli space where they can be understood as
orientifolds of type II theory.  At generic points in the moduli space,
(even those corresponding to infinitesimal deformations away from the
orientifold limit!) the orientifold cannot be described in such a simple
fashion.  We will discuss how one can describe these using M-theory and
F-theory on certain manifolds  which we refer to as U-manifolds
following \cite{alok} because the U-duality group is realised as a
geometric symmetry on these manifolds. 
We shall follow the notation ${\cal
I}_n$ to indicate inversion in $n-$directions, $\Omega$ refers to the
worldsheet parity operation in string theory and $F_L$ is the spacetime fermion
number of the left movers in string theory. 

\subsection{IIA on $S^1/(-)^{F_L}\cdot\Omega\cdot {\cal I}_1$}

Type IIA with string coupling $\lambda^A_{10}$ is constructed by compactifying
M-theory on a circle of radius $R_1$.  The IIA
parameters are given by $\lambda_{10}^A = R_1^{3/2}$ and the ten dimensional
Planck length $l_{10}^A = R_1^{-1/2}$ in units where $l_{11}=1$. 
Compactifying IIA on a circle of radius $R_1^A$ (in ten dimensional units)
is represented by compactifying M-theory on a second circle of radius
$R_2$(in eleven dimensional units).  The two radii
are related by $R_1^A=R_2 R_1^{1/2}$.  One can obtain the following map
which relates operations in
 IIA theory to those in M-theory by studying the action of the fields in the
two theories. ($C$ is the 3-form gauge field of eleven dimensional
supergravity)\footnote{$\Omega$ is not a symmetry of IIA string theory
but $\Omega\cdot {\cal I}_1$ is a symmetry. There seem to be two choices
for the operation $\Omega$  in M-theory: $X^1 \rightarrow -X^1$ or
$C\rightarrow -C$, both of which map $\Omega\cdot {\cal I}_1$ to
symmetries of M-theory. We believe that the two  choices are
continuously connected in the moduli space of the U-manifolds and thus
are equivalent.}

\begin{center}
\begin{tabular}{|c|c|}\hline
IIA operation & M-theory operation \\ \hline
$(-)^{F_L}$ & $C\rightarrow-C$ and $X^1\rightarrow -X^1$\\ \hline
$\Omega$ & $X^1\rightarrow -X^1$\\ \hline
$Y^m \rightarrow -Y^m$ & $X^{m+1} \rightarrow -X^{m+1}$ \\ \hline
\end{tabular}
\end{center}
(Note that in M-theory, in order make ${\cal I}_n$ (for odd $n$)  a
symmetry, one has to take $C\rightarrow -C$.  In the following this will
be assumed even if we do not explicitly indicate it.) We follow the
convention that the coordinates of the IIA theory will be represented by
$Y^m$ (with radius $R^A_m$ in units where the ten dimensional Planck length
$l^A_{10}=1$, if compactified) and those of M-theory by $X^M$ (with radius
$R_M$, if compactified).  

The map discussed above  will enable us to construct models dual to
IIA compactifications in M-theory.  Using the table, 
one can see that the orientifold of IIA on
$S^1/(-)^{F_L}\cdot\Omega\cdot {\cal I}_1$ is given by M-theory on $S^1
\times (S^1/{\cal I}_1)$ in the limit when $R_1 \ll R_2$.  To make this IIA
compactification consistent, one has to include 8 D8-branes at each of the
two fixed points to cancel RR charge which is localised at the fixed
points of the orientifolding group.  Each of the fixed points provides a
$SO(16)$ gauge group leading to an enhanced gauge symmetry of $SO(16)\times
SO(16)$ at the orientifold point.

In M-theory, one can consider another limit which is given by $R_1\gg R_2$, 
which corresponds to the Heterotic string on $S^1$ with parameters
\begin{eqnarray} 
R^H &=& R_1 R_2^{1/2}\quad,\\
\lambda_{10}^{H} &=&R_2^{3/2}\quad, \\
l_{10}^H &=& 1/R_2^{1/2}\quad,
\end{eqnarray} 
where $R_H$ is the radius of the $S^1$, $\lambda_{10}^{H}$ is the ten
dimensional heterotic string coupling and $l_{10}^H$ is the ten dimensional
Planck length.  The nine dimensional string coupling is given by
$\lambda_{9}^H = R_2^{5/4}/R_1^{1/2}$.  The relations between the paramenters 
of the Heterotic string and the IIA orientifold that we obtain here
are consistent with the relations mentioned in ref. \cite{polwit}. This
is an independent check of the map between IIA operations and M-theory
operations which we derived earlier.

As is clear, the above correspondence is at the orientifold limit.
One however needs to understand how to describe things away from the
orientifold limit. For example, one needs to understand the mechanism of
enhancement of gauge symmetry here in IIA language. As we saw, one
needed to include 16 D8-branes for cancellation of $-8$ units of RR
charge at the orientifold fixed points. Using S and T-duality, Polchinski and
Witten have discussed some aspects of the situation away from the
orientifold limit. The important result they obtain is that dilaton
picks up a dependence on the coordinate $X^2$ (in our notation)
in regions between the D8-branes after they have moved away from the
orientifold plane\cite{polwit}. 
Seiberg studied this situation by considering the
worldvolume theory of a probe D4-brane and observed that enhanced gauge
symmetries which involved the exceptional groups $E_n$ occurred at
points where the string coupling diverged\cite{seiberg}. Other groups could be
explained by the coincidence of D8 branes either away from the
orientifold plane ($A_n$ case) or at the orientifold plane ($D_n$ case).
Subsequently it was shown that non-perturbative effects split the
orientifold plane into two, releasing an extra 8-brane!\cite{follow}

All this intricate structure is missing is a geometric picture where the
Narain moduli space is visible and duality is manifest. To achieve this 
we propose that there exists a two dimensional U-manifold
${\cal Y}_1$ which has the following properties:\footnote{All D-branes
of type IIA string theory except the D8-brane can be understood directly
in eleven dimensional supergravity. A related issue is that 
(at the moment) massive IIA
supergravity cannot be derived from eleven dimensional supergravity.
These issues need to be understood in order to understand quantitative
features of ${\cal Y}_1$\cite{massive}.}
\begin{enumerate}
\item ${\cal Y}_1$ is a fibred manifold with fibre $S^1$ and base
$S^1/{\cal I}_1$.
\item The moduli space associated with this manifold is
$$
{{SO(17,1)}\over{SO(17)\times SO(1)}} \times \BR^+ \quad. 
$$
\item There are points in the moduli space where a certain number of
2-cycles (with intersection matrix given by the ADE Cartan matrix)
shrink to zero size. (We expect that this is related to the coalescing of 
degenerate fibres.)
\item There is a point in the moduli space corresponding to constant
fibre where at the two end-points of the base, there are  shrinking
two cycles with the intersection matrix given by the $D_8$  Cartan
matrix.
\end{enumerate}
Assuming that ${\cal Y}_1$ exists, then we would like to claim that
M-theory compactified on ${\cal Y}_1$ should represent the generic
situation  away from the orientifold limit. We identify points where
the fibre degenerates with the location of the D8-branes.
Enhanced gauge symmetries correspond to the massless particles which arise when 
M2-branes wrap around vanishing
two-cycles of ${\cal Y}_1$.\footnote{As this manuscript was being
prepared, a paper by Ashoke Sen has appeared which discusses a similar
mechanism for enhancement of gauge symmetry 
for the orientifold  $T^3/(-)^{F_L}\cdot\Omega\cdot {\cal I}_3$\cite{new}.}
Intuitively, it is easy to see that in the bulk the only possible
intersection matrix is of the $A_n$-type. For example, consider the
situation where two degenerate fibres coalesce
on the base away from the end-points. This corresponds to the $A_1$ case where 
a single two-cycle shrinks to zero.  This argument shows that
$D_n$ and $E_n$ situations cannot arise in the bulk. This
seems to agree with the discussion in ref. \cite{seiberg}. However,
beyond this qualitative picture we do not have a quantitive
correspondence with the details presented in ref. \cite{polwit}. 

Let us summarise the properties of the function which describes
the radius of the fibre circle on ${\cal Y}_1$. We caution the reader
that this has not been derived but guessed by using our picture
of symmetry enhancement and the analysis of refs.
\cite{polwit,seiberg,follow}.
\begin{enumerate}
\item It possesses 18 zeros corresponding to the positions of the
16 perturbative 8-branes and the 2 non-perturbative ones.
\item   It is  a continuous function with cusps at each of the zeros
with the discontinuity in the first derivative being related to the 8-brane
charge.
\end{enumerate}

Is there a candidate for ${\cal Y}_1$? Interestingly, there is one which
was discussed by Morrison and Vafa\cite{morvafa}. We have not checked that it
satisfies all our conditions and will leave it for the future since this
is not the main focus of this paper. The candidate is a real version of
elliptic K3 discussed in the appendix. Consider the manifold
\begin{equation}
y^2 = x^3 + f(z)\ x + g(z)\quad.
\end{equation}
where $f(z)$ and $g(z)$ are homogeneous polynomials (in $z$) of degrees
eight and twelve respectively. Here $x,y,z$ are considered to be real
numbers (unlike in the appendix where they are complex). The equation is
considered to be the equation of a circle with $z$ as the coordinate on the
base.  The number of
parameters which describe this is 18 (22 parameters fix $f$ and $g$ and
one has to subtract three to account for  a global $SL(2,\BR)$ acting on $z$ 
and one for an overall rescaling.). So this passes the simple test that it
gets the dimension of the moduli space correctly. Let us assume
that Tate's algorithm can be applied in this case. Since in the 
orientifold limit of ${\cal Y}_1$, we
expect an $SO(16)$ singularity at the ends, Tate's algorithm implies
that the discriminant has  zeros of order ten at the two end-points of
the base.  The ten zeros can be identified with the nine 8-branes and
the orientifold 8-plane. Again, this seems consistent with picture
we mentioned earlier.

\subsection{IIB on $T^2/(-)^{F_L}\cdot\Omega\cdot {\cal
I}_2$}\label{ttwo}

Following Aspinwall and Schwarz\cite{aspsch}, we construct IIB on $S^1$ in
M-theory as follows.  Consider M-theory compactified on a two-torus with
sides $R_1$ and $R_2$ respectively.  In the limit $R_1 \ll R_2$ and $R_1,R_2
\rightarrow 0$, one obtains type IIB on $S^1$ with length $1/(R_1 R_2)$ in
units where $l_{11}=1$ ($l^B_{10} = 1/R_1^{1\over2}$).  Converting to units
where $l^B_{10}=1$, we get the radius of the $S^1$ to be
$R^B_{1}=1/(R_1^{1\over2} R_2)$. 

The symmetries of IIB theory on $S^1$ can now be seen as symmetries of
M-theory on $T^2$.  The map obtained by studying the action on the fields of
eleven dimensional supergravity is 
\begin{center}
\begin{tabular}{|c|c|}\hline
IIB operation & M-theory operation \\ \hline
$(-)^{F_L}$&  $C\rightarrow -C$ and
$X^{1}\rightarrow -X^{1}$ \\ \hline
$\Omega$& $ C\rightarrow -C$ and
$X^{2}\rightarrow -X^{2}$ \\ \hline
$SL(2,\BZ)$& Mapping class group of the M-theory torus. \\ \hline
$Y^1 \rightarrow -Y^1$& $ C\rightarrow -C$, $X^{2}\rightarrow -X^{2}$ and
$X^{1}\rightarrow -X^{1}$.\\ \hline
$Y^m \rightarrow -Y^m$, $m\neq1$& $X^{m+1}\rightarrow -X^{m+1}$ \\
\hline
\end{tabular}\\
\end{center}
We follow the convention that the coordinates of the IIB theory will be
represented by $Y^m$ (with radius $R^B_m$ in units where the ten
dimensional Planck length $l^B_{10}=1$, if compactified) and those of
M-theory by $X^M$ (with radius $R_M$, if compactified). This will enable
us to construct models dual to IIB compactifications in M-theory.
For example, IIB compactified on an orbifold K3$=T^4/{\cal I}_4$ can be 
realised in M-theory as a compactification on $T^5/{\cal I}_5$ which
is in agreement with known dualities\cite{kth}. 

Thus, type IIB compactified on a two-torus of sides $R^B_{1}$ and
$R^B_{2}$ (in units where $l^B_{10}=1$) can be obtained from M-theory
compactified on a three torus with sides $R_1$, $R_2$ and $R_3$
(in units where $l_{11}=1$). The parameters are related as follows
\begin{eqnarray}
R^B_{1} &=& 1/({R_1^{1/2}R_2}) \nonumber \\
R^B_{2} &=& R_1^{1/2}R_3 \\
\tau_2 &\equiv& {\rm Im}(\tau)= R_2 / R_1\nonumber 
\end{eqnarray}
Note that we have chosen a simple situation where all the tori are rectangular.

We are interested in constructing the orientifold: 
type IIB on $T^2/(-)^{F_L}\cdot\Omega\cdot
{\cal I}_2$. Using the map relating operations in  IIB to those in M-theory
given above, we see that $(-)^{F_L}\cdot\Omega\cdot 
{\cal I}_2$ maps to an inversion of $X^8$ (i.e., the coordinate
associated with the radius $R_3$ in the M-theory compactification).
Thus, we see that the orientifold is obtained  from
M-theory compactified on $T^2 \times (S^1/{\BZ_2})$ in the limit
where the $T^2$ has vanishing area. 

This theory has another limit given by $R_3 \rightarrow 0$ which corresponds to
the $E_8 \times E_8$ heterotic string on $T^2$ with Wilson lines breaking the 
gauge group to $SO(8)^4$. In this limit, 
the ten-dimensional heterotic coupling $\lambda^H_{10}=R_3^{3/2}$ and
the ten-dimensional Planck length $l^H_{10}=1/R_3^{1/2}$.
Now, one can relate the parameters of the two theories using M-theory.  We
obtain
\begin{equation}
 V^B_{T2}\ \tau_2^{1/2}= \lambda_8^H\quad,
\end{equation}
where $V^B_{T2} = R^B_{1}R^B_{2}$ is the area of the IIB torus. Further, 
the complex structure of the
heterotic torus is clearly the  modular parameter of the IIB theory. The complex
structure of the IIB torus 
$$
\rho_2 \equiv {\rm Im}(\rho)=R^B_{2}/R^B_{1}  = R_1 R_2 R_3 = V^H_{T2}\quad,
$$
where $V^H_{T2}$ is the volume of the heterotic torus (in units where
$l_{10}^H =1$). Thus, the complexified K\"ahler structure of
the heterotic string theory is mapped to the complex structure of the
IIB torus. All the relations we have derived are consistent with
the results of Sen mentioned in the appendix. 

As in the type IIA situation, one would like to be able to describe the
situations away from the orientifold limit. The U-manifold ${\cal Y}_2$ 
for this was proposed by Vafa as the first example of an F-theory
compactification\cite{vafa}. ${\cal Y}_2$ here is an elliptic K3 with 
enhancement
of gauge symmetry corresponding to ADE singularities on the U-manifold.
The important issue is that from the IIB viewpoint, there are regions
which involve strong coupling as was shown in a beautiful paper by
Ashoke Sen\cite{ashoke}. Subsequently, there was a nice reinterpretation
of Sen's results using the worldvolume theory of a D3-brane
probe\cite{redo}.  An important result from this analysis is as
follows: At the orientifold point, one has four $D_4$ singularities
which can be understood as coming from four coincident D7 branes  at
the orientifold plane. Non-perturbative effects (from the IIB viewpoint)
split the orientifold plane into two (just as in the IIA case discussed
earlier)! Enhanced gauge symmetry occurs when two-cycles of ${\cal Y}_2$
shrink to zero size with the gauge group given by the intersection
matrix.  Some relevant details are discussed in the appendix.

\subsection{IIA on $T^3/(-)^{F_L}\cdot\Omega\cdot {\cal I}_3$}

We include the case of IIA on $T^3/(-)^{F_L}\cdot\Omega\cdot {\cal I}_3$
to include another example in our list.
This orientifold can be mapped to M-theory on $S^1\times (T^3/{\cal
I}_3)$. The enhanced gauge symmetry here is $SO(4)^8=SU(2)^{16}$.
One can again ask what happens when one goes away from the
orientifold limit. Equivalently, we can ask if we can find a four
dimensional fibred U-manifold ${\cal Y}_3$ with fibre $S^1$ which describes the 
situation of varying IIA coupling. The answer here is that ${\cal Y}_3$
is a K3. One valid objection to this is that a generic K3 is not
a fibred manifold with fibre $S^1$. However, it is possible that
in the complete moduli space of K3, there is a subspace (of codimension
zero) which satisfies this criterion. There is a recent paper by
Aspinwall which might be relevant in understanding this\cite{asp}.
Further, this is consistent with the known duality of the Heterotic
string on $T^3$ and M-theory on $K3$.

\subsection{A summary}

For the reader who is not interested in the technical details,
we now present a summary of the results which we have discussed in the previous 
subsections. We provided maps which directly relate
orientifolds of type II string theory to the $E_8\times E_8$ heterotic
string at special points in their moduli spaces using M-theory. These
maps can also be obtained by a complex sequence of T and S duality
operations involving the type I and the two Heterotic strings. (An
example due to Sen is provided in the appendix. A similar sequence has also been
discussed by Polchinski and Witten\cite{polwit}.) Away from the orientifold 
limits, these orientifolds are best understood as compactifications of M- or
F- theory on U-manifolds ${\cal Y}_d$. One common feature which emerges in the
examples considered is that enhancement of gauge symmetry can be
understood in these theories as coming from the wrapping of
the M2-brane on shrinking two-cycles of the U-manifold. From the type II
picture this is understood as coming from the coinciding of D-branes.
As we will see later, in M(atrix) theory the first viewpoint is precisely the
mechanism by which enhanced gauge symmetry and (not surprisingly)
the twisted sector fermions emerge.

\section{Implications for M(atrix) theory}

In the previous section, using U-manifolds, we saw how one could describe 
situations corresponding to deformations away from orientifold limits
inspite of the fact that the string theory could be at strong coupling. In
M-theory we achieved this by considering a fibred manifold with fibre an
$S^1$ whose radius was related to the string coupling constant of IIA
theory. In F-theory, this was achieved by considering another fibred
manifold with fibre $T^2$ with the complex structure of the fibre torus
being the dilaton-RR scalar of IIB string theory. Enhanced gauge
symmetry is related to the  vanishing of certain cycles on the U-manifold
and the moduli space is recovered as the moduli space of the
U-manifold. 

In this section, we will discuss the implications of these observations for
M(atrix) theory. We discuss the case of Heterotic M(atrix) theory
on $T^2$ and then discuss the cases of $S^1$ and $T^3$.
The important result which we will achieve is that U-duality will be
manifest and further the twisted sector (``fermions'') appear naturally
without being postulated.

The basic idea is that in Heterotic M(atrix) theory, moving away from
the orbifold limit corresponds to letting the coupling constant of the
SYM vary with the base such that the appropriate amount of supersymmetry
is preserved. Additional constraints come from anomaly cancellation.
(Issues related to this have been discussed by Ho\v rava\cite{petr} and
by Kabat and Rey\cite{krey}.)
Further, there can be regions where the coupling constant
might be strong. This might seem to be a rather difficult problem to
solve. 
Actually, this is rather similar to what we observed for the case of
the type II orientifolds we considered in Sec. \ref{orient}. 
So, as in that case, we propose to consider an auxiliary manifold
whose fibre is the coupling constant of the SYM. For Heterotic
string theory on $T^d$ the auxiliary manifold ${\cal Y}_d$
has to satisfy the following conditions:
\begin{enumerate}
\item ${\cal Y}_d$ is a fibred manifold with fibre $S^1$ (for odd $d$)
and $T^2$ (for even $d$). (This is related to the fact that in four
dimensions one can complexify the coupling constant of SYM by including
the $\theta$ - term $F\wedge F$.) 
\item The moduli space associated with this manifold is
$$
 {{SO(16+d,d)}\over{SO(16+d)\times SO(d)}}  \times \BR^+\quad.
$$
\item There are points in the moduli space where certain number of
2-cycles (with intersection matrix given by the ADE Cartan matrix)
shrink to zero size.
\item There is a point in the moduli space corresponding to constant
fibre where the base of the manifold is $T^d/{\cal I}_d$ with
two cycles shrinking to zero size at the fixed points whose
intersection matrix is given by the $SO(32/r)$ Cartan matrix where $r=2^d$.
\item Enhanced gauged symmetry should occur at precisely the same
point where the dual type II compactification has gauge symmetry
enhancement.
\end{enumerate}

Following the discussion in Sec. \ref{orient}, the U-manifolds
we described for the type II orientifolds seem the perfect candidates. 
However, one has to decide on which theory these manifolds have to be
compactified. The answer is fixed by the condition that for large base
(small fibre) and away from possible singular fibres, the theory should look 
like SYM on $T^d$ (for $d<4$). A natural candidate which satisfies this
is the theory  which describes M(atrix) theory on $T^s$ for some $s>d$.
This leads us to  propose the following as a replacement of the SYM 
prescription.
\begin{center}
{\it Compactify  M(atrix)   theory  on  $S^1\times {\cal Y}_d\quad,$ }
\end{center}
where the theory to be chosen is decided using the criterion mentioned
above.  This can be considered as the {\it main result} of this paper.
For example, for ${\cal N}=8$ SYM in
2+1 dimensions is replaced by ${\cal N}=4$ U(N) SYM in 3+1 dimensions 
(this theory describes M(atrix) theory on $T^3$).

Immediate consequences are: U-duality is manifest since it is encoded in
the geometry of ${\cal Y}_d$. 
Unlike in the case of the orientifolds, we however seem to
have an extra modulus associated with the size of the $S^1$.
The resolution is that in all the
examples that we consider, the theory is superconformal and hence one
of the scales is not a modulus. The proposed theory is non-anomalous
since we expect compactification to preserve the non-anomalousness
of M(atrix) theory on $T^s$. Thus, these are simple consistency checks
which our proposal passes. 
We shall now consider the proposal for three cases $d=1,2,3$. We first 
discuss the case of $d=2$ since many properties and issues can be 
made explicit.

\subsection{Heterotic M(atrix) theory on $T^2$}

In M(atrix) theory, the heterotic string on $T^2$ is described by the
orbifold SYM theory on $\widetilde{S}^1 \times \widetilde{T}^2/\BZ_2$.
Anomaly considerations requires one to include 32 fermions in the theory
as a twisted sector. Distributing the fermions
equally among the four fixed circles leads  to an $SO(8)^4$ gauge symmetry 
in the target space. This is precisely the same gauge group which arose in the
IIB orientifold we considered and was the motivation for the special
limit for F-theory considered in the appendix. 

From Sec. \ref{orient}, we see that  ${\cal Y}_2$ is an elliptic K3 and 
we should consider the M(atrix) theory describing  M-theory on $T^4$. This
is a theory with (2,0) supersymmetry in 5+1
dimensions which on compactifying on a
two torus reduces to SYM as mentioned earlier\cite{five}. 
So our proposal implies that Heterotic M(atrix)
theory on $T^2$ is described by {\it compactifying the (2,0) theory
on $S^1\times {\cal Y}_2$}. It clearly satisfies all the conditions
we described. Among them it satisfies the constraint that it possess a
limit with constant fibre such that one has $SO(8)^4$ enhanced gauge
symmetry with base $T^2/\BZ_2$. As shown in the appendix this limit
exists. Berkooz and Rozali have also considered compactifying the
(2,0) theory on an elliptic K3 in relation to F-theory on K3\cite{BZ}.
However, they attempted to relate to the heterotic string at the point where
the enhanced gauge symmetry is $E_8\times E_8$. Thus it was not possible to
relate to the existing prescriptions which lead to $SO(8)^4$. Our
proposal is however consistent with their result.

Another important consequence is that the twisted sector fermions
(which are equivalent to chiral bosons in 1+1 dimensions) can be seen 
clearly. At the orientifold limit, one has four two-cycles shrinking to
zero with their intersection matrix given by the $D_4$ Cartan matrix.
For the U(1) case, the (2,0) theory is a theory of a single tensor
multiplet which has a self-dual two-form gauge field. Away from the
singularities, this provides SYM as described by
Verlinde\cite{verlinde}. At the singularities, the same two-form gauge
field provides four chiral bosons which on fermionising transform in
the 8 of $D_4$. We thus recover the twisted sector fermions which had
to be postulated at the oribifold limit. The case for U(N) needs a better
understanding of the (2,0) theory but it is not unlikely that the bosons
will transform in the appropriate representation. 

Let us now attempt to return to the SYM picture. When the base of
${\cal Y}_2$  (which is a $\BC\BP^1$) is large, the geometry of ${\cal
Y}_2$ dictates the variation of the SYM coupling constant $\rho =
{\theta\over {2\pi}} + {{4\pi i}\over{g^2}}$. An important condition is that
$\rho$  varies holomorphically with respect to the complex
coordinate of the base $\BC\BP^1$. 
Thus contrary to the
naive guess that the base manifold of the SYM is $\widetilde{T}^2/{\cal
I}_2\times S^1$, it is given by $\BC\BP^1\times S^1$. At 24 points
(corresponding to the zeros of the discriminant of the elliptic K3),
the SYM is at strong coupling. Enhanced gauge symmetries 
correspond to the coalescing of zeros with the gauge
group given by Tate's algorithm\cite{morvafa}. In the SYM picture, this 
corresponds to a global symmetry acting on chiral bosons (fermions)
living on the $S^1$ located at the zero of the discriminant.  
For example, an $E_8$ symmetry occurs when ten zeros of the discriminant
coincide with the order of the zeros of $f$ is larger than three  and that
of $g$ is equal to five at the same location. Thus in the moduli space,
there is a point where $E_8\times E_8$ occurs\cite{morvafa}.
We interpret the condition that $\rho$
depends holomorphically on the coordinate of the $\BC\BP^1$
as a BPS condition which breaks half of the ${\cal N}=4$
supersymmetry. It is of interest to derive this from first principles in
3+1 SYM. We hope to report on some of these aspects in the future.

\subsection{Heterotic M(atrix) theory on $S^1$}

Using an approach different from ours, Kabat and Rey have considered this
problem directly from the SYM in 2+1 dimensions. We thus will be able
to compare our proposal to their results.
Here ${\cal Y}_1$ is the two dimensional U-manifold which we discussed
in Sec. \ref{orient}. Our proposal suggests that we compactify
3+1 dimensional U(N) SYM  on $S^1\times {\cal Y}_1$. When the base of
${\cal Y}_1$ is large and away from singular fibres, 
locally the theory reduces to SYM in 2+1
dimensions using standard dimensional reduction. 
Further, at the limit where ${\cal Y}_1$ has constant fibre,
at the endpoints of the base, one has eight 2-cycles whose intersection
matrix is given by the $D_8$ Dynkin diagram, the field strength of the
gauge field can provide scalars living on the 1+1 dimensional
circle. We do not understand this mechanism very well and thus this
should only viewed as a possible scenario. We do not see how the
chirality of the bosons emerge.

We now  comment on the relationship to the recent work of Kabat
and Rey\cite{krey}. Since our proposal  leads to a varying coupling
constant it is similar to their work. However, since we do not fully
understand ${\cal Y}_1$ as yet, we cannot make a precise comparision.
Kabat and Rey point out that T-duality is realised as S-duality in
Heterotic M(atrix) theory. This can probably be understood  in our proposal 
as a part of the electric-magnetic duality of SYM in 3+1 dimensions
thus providing a different perspective to their result.
(This might also be related to the
results of Susskind and Ganor et al.\cite{sdual}, who showed that T-duality 
in M(atrix) theory on $T^3$ is realised as S-duality of 3+1 dimensional
SYM.) 

Kabat and Rey obtain a potential for the scalars and also find different
behaviour for the centre of mass U(1) and the SU(N) coupling constants.
We do not understand how this feature will emerge in our situation. This
might need a better understanding of the of ${\cal Y}_1$ and details
of how the dimensional reduction works. However, since in the analysis of
Kabat and Rey, this follows from rather general considerations, we
expect that such behaviour should emerge.

The only disagreement we have with their result is the choice
of the variation for the coupling constant of the SYM (which is related to 
z(y) in their notation). 
Even though we cannot make a precise statement 
about the functional form of this function since we do not know the
detailed structure of ${\cal Y}_1$, our mechanism of gauge symmetry
enhancement
provides insight into the form of the variation of coupling constant.
However, the crucial features which are required in the variation
for  the coupling constant as follows from the analysis on Kabat and Rey seem
to be present. Thus we do not consider it to be a major disagreement
with their analysis.

We will now discuss the feature of the variation of the coupling 
constant in our proposal. In our proposal, the radius of the fibre
is related to the coupling constant of SYM (and string coupling in
the IIA orientifold).  At a zero, SYM is a strong coupling since the
2+1 dimensional coupling constant is related to the inverse of the
radius ($g^2 \sim 1/R$, assuming standard dimensional
reduction).  Consider the $A_1$ situation which occurs when two degenerate
fibres coalesce.   Since the radius of the fibre is positive definite, 
it  will have to increase away from the zero but at some point
it should turn around to touch zero again. This is the picture forced on
us by the occurance of two neighbouring zeros. We do not see this
behaviour in the variation of the coupling as chosen by Kabat and Rey.
However, there  must be a neighbourhood of near a  zero where
one can approximate the function locally by a linear function such that
locally it agrees with the form of Kabat and Rey.
In the analysis of Kabat and Rey, the jump in the derivative was
related to the Chern-Simons coefficient and thus this feature is captured
in our scenario. A more detailed comparision will require understanding
${\cal Y}_1$ as well as the issues related to D8-branes and massive IIA
supergravity being obtained from eleven dimensional
supergravity\cite{massive}.

\subsection{Heterotic M(atrix) theory on $T^3$}

Here ${\cal Y}_3$ is a K3 and hence our proposal would correspond to
considering the (2,0) theory on $S^1 \times K3$ which is in agreement
with an existing proposal\cite{seven,BZ,hrey}.\footnote{This example
is rather different from  the first two examples we just described. 
This theory is not expected to be described by SYM even in its orbifold
limit. Thus what we are doing  
corresponds to an extrapolation of the previous examples.
The only justification which we can provide is that
the result seems consistent.} From the arguments
given in sec. \ref{orient}, Heterotic theory will be visible only
in the subspace in the moduli space of K3 where it is a fibred manifold
with fibre $S^1$. This is in agreement with related remarks of Berkooz
and Rozali\cite{BZ}. The mechanism of gauge symmetry enhancement here is
similar to the case of  Heterotic M(atrix) theory on $T^2$.

\section{Conclusion}

In this paper, we have discussed Heterotic M(atrix) theory on $T^d$ at
generic points in their moduli spaces. We have proposed certain
U-manifolds ${\cal Y}_d$ which have the property that M(atrix) theory
compactified on $S^1\times {\cal Y}_d$ describes
Heterotic M(atrix) theory on $T^d$ at generic points in its moduli space. 
Evidence for our
proposal is provided by manifest U-duality, a limit where we recover the
orbifold SYM on $S^1\times T^d/\BZ_2$ with the correct gauge symmetry
and a natural mechanism for the appearance of the postulated fermions
precisely at the point where enhanced gauge symmetry occurs (we do not
understand this too well for the case of $d=1$). In conjunction with the
results of Kabat and Rey, we hope that this is a step towards
understanding Heterotic M(atrix) theory in more detail.
One aspect which is lacking here is some kind of first principles
derivation of (atleast) some of the features starting from the orbifold SYM. 
This is being currently studied\cite{current}. 

One striking feature which seems to come out of this proposal is the
disappearance of the spacetime picture. The $T^2$ case is rather
striking since the dual torus gets replaced by a sphere. 
This is similar to what
happens in M(atrix) theory on $T^4$ and $T^5$ and must be a generic
situation when U-duality is manifest\cite{four,five}. Spacetime can be only 
recovered in some special limits as emphasised by these authors. 

The proposal of using U-manifolds is more general than the examples we
considered in this paper. It might prove useful in situations with even
fewer unbroken symmetries. An interesting example is that of the
Heterotic string compactified on K3. Now one has to specify additional
data corresponding to a choice of vector bundle on K3. This is dual to
F-theory on certain elliptic CY 3-folds\cite{morvafa}. A natural guess
for describing these cases in M(atrix) theory is to consider M(atrix) theory 
on the same CY$\times S^1$. However this is a seven dimensional base space
and hence  we need to understand M(atrix) theory 
on $T^6$ and maybe even $T^7$. M(atrix) theory in this case might provide 
another window
into understanding the relation between vector bundles on K3 which occur
on the heterotic side and the CY 3-folds which occur in the F-theory
description. We also believe that our results might be relevant to the
understanding of D-branes in curved space\cite{curved}.

\noindent {\bf Acknowledgements:} This work is supported by a fellowship
from the Alexander von Humboldt Foundation. We are grateful to Werner
Nahm and the rest of the Theory Group at Bonn for their kind hospitality.
We thank F. Rohsiepe, R. Schimmrigk, A. Sen and S. Sethi for useful
conversations and S.-J. Rey for useful comments on an
earlier version of the manuscript. 
 
\appendix
\begin{flushleft}
{\Large \bf Appendix}
\end{flushleft}
\section*{F-theory on elliptic K3}

We collect some facts about F-theory on an elliptic K3 which are
relevant.  An elliptic K3 surface is given by the equation
\begin{equation}
y^2 = x^3 + f(z)\ x + g(z)\quad.
\end{equation}
where $f(z)$ and $g(z)$ are homogeneous polynomials (in $z$) of degrees
eight and twelve respectively.
The K3 is the fibred manifold with base $\CP^1$ (and coordinate $z$)
and the fibre is a torus described by the equation given above. The modular
parameter of the torus is implicitly given by
\begin{equation}
j(\tau(z)) = {{4 (24 f)^3}\over {27g^2 + 4f^3}}\quad,
\end{equation}
where $j$ is the standard j-function. Generically, the fibre degenerates
at the 24 zeros of the discriminant $\Delta \equiv 4 f^3 + 27 g^2$.

Since we will be interested in understanding the relationship of
F-theory compactified on elliptic K3 to heterotic strings compactified on $T^2$
in M(atrix) theory, it is of interest to consider a special limit. This
is a limit where the modular parameter $\tau(z)$ is constant over the
base. Several possibilities exist\cite{ashoke,sunil,morvafa}. We shall
however consider the one considered by Sen\cite{ashoke}. This
corresponds to the case where the discriminant has four zeros, each of
order six.\footnote{The fact that $\Delta$ has zeros of order six at 
each $z_i$ is important. Deforming away from the orientifold limit,
which in the elliptic K3 picture corresponds to separating the zeros,
one obtains six separate zeros. Four of them can be identified with the
positions of the D7-branes. Sen has argued that the other two
zeros corresponds to a splitting of the orientifold plane or
equivalently the emitting of a 7-brane by the orientifold plane.}
 
Explicitly, this corresponds to choosing 
\begin{equation}
g=\phi^3\ {\rm and}\ f=\alpha\ \phi^2\quad,
\label{elim}
\end{equation}
where $\phi=\prod_{i=1}^4 (z-z_i)$. 
This leads to an enhanced gauge symmetry of
$SO(8)^4$ in the F-theory compactification. 
In addition, the base can now
be considered to be a torus whose complex modulus `$\rho$' corresponds to the
cross-ratio of the locations of the four zeros of the discriminant. 
The moduli of this theory are now given by $\tau$, $\rho$ and the
size of the base $V^B_{T2}$. At this special limit, F-theory on K3 can be
identified with type IIB string on $T^2/(-)^{F_L}\cdot\Omega\cdot {\cal I}_2$. 

We shall now use a chain of dualities to map this F-theory
compactification to a compactification of the $E_8\times E_8$ Heterotic
string on a two-torus. This will enable us to obtain an explicit
relationship of the moduli on both sides. 
First, we shall T-dualise on both circles of the
base. This maps the $\BZ_2$ transformation $(-)^{F_L}\cdot\Omega\cdot
{\cal I}_2$ to the transformation $\Omega$. Thus we can map the F-theory
compactification to type I on the dual torus. We can now use the
type I - Heterotic duality to obtain a map to the 
$SO(32)$ Heterotic string. Finally, by T-dualising on one of the
circles, we obtain a map from the F-theory compactification to the
$E_8\times E_8$ Heterotic string. The moduli on the heterotic side
given by the complex and K\"ahler moduli of the torus (on which the heterotic
string is compactified) get mapped to the IIB moduli $\tau$ and $\rho$
respectively. Further, the eight-dimensional heterotic string coupling
$\lambda^H_8$ is related to the IIB moduli as
\begin{equation}
\lambda^H_8 = V^B_{T2}\ \tau_2^{1\over2}\quad.
\end{equation}
The above chain of dualities as well as relations just mentioned have
been described by Sen\cite{ashoke}.

\end{document}